# Layer and doping tunable ferromagnetic order in two-dimensional CrS$_2$ layers

Cong Wang[1], Xieyu Zhou[1], Yuhao Pan[1], Jingsi Qiao[1], Xianghua Kong[1], Chao-Cheng Kaun[2], Wei Ji[1,*]

[1]*Beijing Key Laboratory of Optoelectronic Functional Materials & Micro-Nano Devices, Department of Physics, Renmin University of China, Beijing 100872, P.R. China*

[2]*Research Center for Applied Sciences, Academia Sinica, Taipei 11529, Taiwan, R. China*
Email: wji@ruc.edu.cn

**Abstract:**

Interlayer coupling is of vital importance for manipulating physical properties, e.g. electronic bandgap, in two-dimensional materials. However, tuning magnetic properties in these materials is yet to be addressed. Here, we found the in-plane magnetic orders of CrS$_2$ mono- and few-layers are tunable between striped antiferromagnetic (sAFM) and ferromagnetic (FM) orders by manipulating charge transfer between Cr $t_{2g}$ and $e_g$ orbitals. Such charge transfer is realizable through interlayer coupling, direct charge doping or substituting S with Cl atoms. In particular, the transferred charge effectively reduces a portion of Cr$^{4+}$ to Cr$^{3+}$, which, together with delocalized S $p$ orbitals and their resulting direct S-S interlayer hopping, enhances the double-exchange mechanism favoring the FM rather than sAFM order. An exceptional interlayer spin-exchange parameter was revealed over -10 meV, an order of magnitude stronger than available results of interlayer magnetic coupling. It addition, the charge doping could tune CrS$_2$ between p- and n-doped magnetic semiconductors. Given these results, several prototype devices were proposed for manipulating magnetic orders using external electric fields or mechanical motion. These results manifest the role of interlayer coupling in modifying magnetic properties of layered materials and shed considerable light on manipulating magnetism in these materials.

# I. INTRODUCTION

Geometric, electronic, vibrational, thermal and optical properties of layered two-dimensional (2D) materials, like graphene[1-4], transition-metal dichalcogenides (TMDs)[5-9] and black phosphorus (BP)[10-15], have received considerable attention in the past decade. Magnetic properties were, however, rarely mentioned and are thus of particular interest. Magnetism in 2D materials could be introduced by dosing adatoms[16-20], atom substitution[21], creating vacancies[16, 22] or boundaries[23] or constructing atomic edges[24, 25]. Nevertheless, the long-range magnetic ordering was argued less stable in 2D because of the largely reduced size in the perpendicular direction according to the Mermin-Wagner theorem[26]. Therefore, the long-range ordering could be easily eliminated by thermal fluctuation, leading a ferromagnetic (FM) or anti-ferromagnetic (AFM) to paramagnetic transition. It was very recently reported that magnetic ordering persists even if the thickness is down to monolayer in $CrI_3$ [27, 28]. Another strategy relies on a magnetic exchange field that stabilizes the long range ordering with magnetic anisotropy in a bilayer $Cr_2Ge_2Te_6$[29]. These two studies compellingly support the existence of long-range magnetic ordering in 2D, which may allow surface sensitive techniques to measure 2D magnets and improve the feasibility of theoretical models built in 2D for real materials under investigation. However, the knowledge of manipulating magnetism in 2D is yet to be revealed.

Interlayer coupling has been manifested to be of paramount importance in manipulating physical properties [4, 13, 14, 30-35] in 2D materials. The layer-tunability is a result of layer-dependent electronic structures caused by strong interlayer electronic coupling, as named covalent-like quasi-bonding (CLQB) [14, 32, 36, 37]. In CLQB-governed 2D layers, the distance between two or more layers was pushed close enough by the van der Waals attraction that the wavefunctions, especially from $p_z$ or $d_{z^2}$ orbitals, of two adjacent layers are forced to overlap forming bonding and anti-bonding states. Such hybridization should substantially modify electron distributions in both layers and thus may change either in-layer or inter-layer magnetic coupling. As evidenced by the layer-dependent interlayer magnetic orderings in $CrI_3$, interlayer coupling might play an essential role in tuning interlayer magnetism. However, its ability of varying *in-plane* magnetic ordering is neither theoretical suggested or experimentally proved. The interlayer engineering [38] aside, charge doping by, e.g. ionic liquid gating, has been becoming a popular route to manipulate electronic properties in 2D. A recent work reported $O^{2-}$ ion doping induced change of magnetism in Co/SrCoO$_x$ thin films[39]. Nevertheless, the O atom itself, rather than the charge solely, plays a key role. It is still lack of an example for a purely electron- or hole-doping governed transition of neither in-layer nor inter-layer magnetic orders in layered materials.

In this article, we report a theoretical investigation on the layer- and doping-dependent magnetism in CrS$_2$ few-layers. Monolayer CrS$_2$ is in a striped anti-ferromagnetic order (sAFM) [40-43]. It undergoes a sAFM to FM transition if one or more additional layers were stacked onto the monolayer. Such transition is a result of a strong interlayer CLQB, which weakens the in-plane Cr-S σ bonds and transfers electrons from $e_g$ to partially occupied $t_{2g}$ orbitals of Cr. It is exceptional that the interlayer spin-exchange coupling parameter of -10.8 meV is nearly an order of magnitude larger than previously reported values in other materials[29, 44, 45]. Layer stacking aside, we also discussed the roles of sole electron or hole doping or elemental substitution[46, 47] in tuning magnetism in 2D. These results suggest CrS$_2$ a FM/sAFM switchable 2D layer and illustrate the ability of interlayer engineering for tuning either in-layer or inter-layer magnetism in layered magnets.

## II. METHODOLOGY

### A. DFT calculations.

Density functional theory calculations were performed using the generalized gradient approximation for the exchange-correlation potential, the projector augmented wave method[48] and a plane-wave basis set as implemented in the Vienna ab-initio simulation package (VASP)[49]. The energy cutoff for plane wave of 700 eV and 600 eV were adopted for calculation of structural relaxation and electronic structures, respectively. A 16 × 10 × 1 *k*-mesh was used to sample the first Brillouin zone of the adopted 1×√3×1 rectangular supercell. A second-order Methfessel-Paxton smearing method (with a sigma value of 0.1 eV) and a Bloch-corrected tetrahedron method were used for structural relaxations and total energy calculations, respectively. Dispersion correction was made at the van der Waals density functional (vdW-DF) level, with the optB86b functional for the exchange potential[50]. All atoms in the supercell were allowed to relax until the residual force per atom was less than 0.01 eV/Å. A 2×4 supercell was adopted to calculate the exchange parameters. All configurations were fully relaxed.

## B. DFT+U method.

The DFT+U method considers orbital dependence of the Coulomb and exchange interactions, which gives a qualitative improvement compared with standard DFT to transition metal systems for either magnetic moments or interatomic exchange parameters[51]. Here, the adopted value was suggested from the literature[52] where a U value of 0.26 Ry and a J value of 0.053 Ry, equivalent to an effective U value of approximate 2.8 eV, was unveiled in systems of Cr impurities embedded in metal. Since $CrS_2$ is metallic, the Coulomb screenings in both cases shall be comparable and the choice of 2.8 eV is thus reasonable for describing the on-site U effect in $CrS_2$.

We have also self-consistently calculate the U value using a linear response method[53]. The calculated U value is in a range from 1.5 eV to 2.5 eV, close to the value reported in the literature (from 2eV to 4eV)[52, 54-57] and the value we adopted (2.8 eV). In addition, we checked several effective U values (Supplementary Fig. S1 [58]) and found it is robust to show the found large interlayer exchange coupling parameter and the doping induced sAFM-FM transition. The choice of 2.8 eV is thus reasonable for describing the on-site U effect in $CrS_2$.

The experimental groundstate of $CrSe_2$, an analogue compound of $CrS_2$, is magnetic[59, 60]. A 2H-non-magnetic groundstate of $CrSe_2$ was unveiled in our calculations without a U term (0.20 eV more stable than the 1T-sAFM state). If we consider the U term, a 1T-sAFM groundstate was compellingly suggested with the U value varying from 1eV to 5 eV (over 0.41 eV more stable for U=2.8 eV). Given the experiment magnetic groundstate, it thus indicates the U term is paramount for obtaining the correct groundstate of Cr chalcogenides using state-of-the-art DFT methods.

The agreement of our results with the experiments aside, a previous calculation [61] also reported the 1T-magnetic groundstate of CrSe$_2$ using a localized spherical wave (LSW) method. This method assumes spherical tight-binding wave functions and fill the space of lattice using occupied or empty spheres and predicted the magnetic groundstate without U. Here, the choices of sphere positions/types and the radii of spheres are all parameters to be determined or fitted. In light of this, we thus have confidence that our DFT+U method is a state-of-the-art method for modelling Cr dichalcogenides.

### C. Carrier mobility estimation.

Phonon-limited carrier mobility in CrS$_2$ few-layer with a finite thickness $W_{eff}$ is expressed as [62-65]:

$$\mu_{film} = \frac{\pi e \hbar^4 C_{film}}{\sqrt{2}(k_B T)^{3/2}(m^*)^{5/2}(E_1^i)^2} F$$

Here, $m^*$ represents the effective mass along the transport direction and $E_1$ is the deformation potential constant of the VBM (hole) or CBM (electron) along the transport direction, which is determined by $E_1^i = \Delta V_i/(\Delta l/l_0)$. Here $\Delta V_i$ is the energy change of the $i$<sup>th</sup> band under proper cell compression and dilatation (by a step 0.5%), $l_0$ is the corresponding lattice constant along the transport direction and $\Delta l$ is the deformation of lattice constant. Variable $C_{film}$ is the elastic modulus of the longitudinal strain in the propagation direction, which is derived by $(E - E_0)/V_0 = C(\Delta l/l_0)^2/2$; $E$ represents the total energy and $V_0$ represents the lattice volume at the equilibrium for 2D systems. A crossover function $F$ bridges the 2D and 3D cases, which is estimated by:

$$F \equiv \frac{\sum_n \left\{ \frac{\sqrt{\pi}}{2} [1 - \text{erf}(\Omega(n))] + \Omega(n) e^{-\Omega^2(n)} \right\}}{\sum_n [1 + \Omega^2(n)] e^{-\Omega^2(n)}}$$

where

$$\Omega(n) \equiv \sqrt{\frac{n^2 \pi^2 \hbar^2}{2 m^* W_{eff}^2 k_B T}}$$

The erf() represents an error function and the summation over integer is due to quantum confinement along the z-direction. Effective thickness of the film ($W_{eff}$) is expressed by:

$$\frac{1}{W_{eff}} = \int_{-\infty}^{+\infty} P_i(z) P_f(z) dz = \sum_n \frac{\rho_i^n(z)}{N \Delta z} \cdot \frac{\rho_i^n(z)}{N \Delta z} \Delta z$$

Here, $P(x)$ is the electron probability density along the z direction. We divided the space along the $x$ direction into $n$ parts by $\Delta z$. Variable $\rho^n(z)$ is the sum of the number of electrons $n^{th}$ region along the $z$ direction. Here, $N$ is the total number of valence electrons in the film, $i$ and $f$ represent equilibrium and deformed films, respectively.

## D. Plotted charge densities

The spin charge density shows the distribution of charge with different spin components. Two sets of data, total charge density (spin up plus spin down) and magnetization density (spin up minus spin down), are available in spin polarized calculations. The charge density with the spin-up (-down) component can be extracted by adding (subtracting) the two sets of data. The atomic differential charge density was plotted to show the charge redistribution before and after Cr and S atoms forming a $CrS_2$ monolayer, which is determined by $\Delta \rho_a = \rho_0 - \rho_a$. Here $\rho_0$ is the charge density converged in an electronic self-consistency loop which contains the interaction between bonded atoms while $\rho_a$ is obtained by a non-self-consistent calculation for summing over atomic charge densities.

## E. Implementation of charge doping

Direct charge doping was applied on S atoms with the ionic potential method [66]

to tune magnetic orders and to understand the role of interlayer couplings. Electrons (holes) are removed from a 2p core level of S and placed into the lowest unoccupied band of CrS$_2$. This method ensures the doped charges being located around the S atom and keeps the neutrality of the layer. We also plotted the differential charge density of doped systems using $\Delta\rho_d = \rho_d - \rho_0$, where $\rho_d$ is the charge density of a doped CrS$_2$.

### F. Interlayer force constant

The whole layer is regarded as one rigid body in a rigid-layer vibrational mode [32, 67]. The interlayer force constants were obtained by summing interatomic force constants over all atoms from each of the two adjacent layers. The matrix of interatomic force constants, essentially the Hessian matrix of the Born–Oppenheimer energy surface, was defined as the energetic response to a distortion of atomic geometry in DFPT[68], which reads as $D_{ij} = \frac{\partial^2 E(R)}{\partial R_i \partial R_j}$. Here, $R$ is the coordinate of each atom and $E(R)$ is the groundstate energy.

## III. RESULTS AND DISCUSSION

### A. Structure and spin-exchange coupling of monolayer CrS$_2$

A striped anti-ferromagnetic state in the 1T-phase [1T-sAFM, see Fig. 1(a) & 1(c)] is the most energetically favored (over 23 meV/Cr) state of a monolayer CrS$_2$, rather than the previously believed non-magnetic 2H-phase [2H-NM, Fig. 1(b) and Supplementary Fig. S2] [40-43]. The sAFM groundstate persists under six different $U$ values (1-5 eV, Supplementary Fig. S1a [58]). The magnetic moment of each Cr atom slightly depends on the adopted magnetic order, in a range from 2.4 – 2.8 μ$_B$. These moments reduce by 0.6 – 0.8 μ$_B$ if the $U$ term was not considered (Supplementary Table S1 [58]). The formation of magnetic stripes reduces the 3-fold structural symmetry that a Jahn-Teller distortion occurs in the sAFM configuration, namely $a$ = 3.31 Å, $b$ = 5.45 Å, $r_1$ = 2.38 Å and $r_2$ = 2.36 Å, $\theta_1$=88.2°, $\theta_2$=84.6°. The smaller $\theta_2$ prefers to weaken Hund's law and thus favoring AFM coupling. In FM [Fig. 1(d)], however, the symmetry maintains and $\theta_2$ approaches 90° (87.7°), resulting in $a$ shrunk by 0.02 Å and $b$ expanded by 0.25 Å.

Spin-exchange coupling parameters were extracted based on a third-nearest Heisenberg model, as follow.

$$H = H_0 + J_1 \sum_{\langle ij \rangle} \vec{S_i} \cdot \vec{S_j} + J_2 \sum_{\langle\langle ij \rangle\rangle} \vec{S_i} \cdot \vec{S_j} + J_3 \sum_{\langle\langle\langle ij \rangle\rangle\rangle} \vec{S_i} \cdot \vec{S_j}$$

Here, $J_1$, $J_2$ and $J_3$ represent the first-, second- and third- nearest couplings, respectively, as illustrated in Fig. 1(a). Given the four configurations shown in Fig. 1(c-f), we derived $J_1$ = 47.7 meV, $J_2$ = -38.9 meV and $J_3$ = 9.6 meV. The larger $|J_1|$ than $|J_2|$ and the positive $J_1$ are consistent with the favored sAFM configuration.

## B. Magnetic coupling mechanism of monolayer CrS$_2$.

Figures 2(a) and 2(b) show the side- and top-views of the spin density of a CrS$_2$ monolayer in the sAFM configuration, which reflects the broken symmetry around S atoms. They also show localized S $p_z$ densities for one spin-component and delocalized $p_x$ or $p_y$ densities for the other spin-component, resulting a nearly cancelled total magnetic moment of 0.034 μ$_B$. Atomic differential-charge-density of the monolayer depicts charge accumulation residing primarily around S atoms but varnishing around Cr atoms [Fig. 2(c)]. However, charge reduction (Fig. 2(d)) occurs mainly in the Cr $d_{x2-y2}$ and $d_{z^2}$ orbitals and partially at the interatomic region of S atoms. An electron transfer is evidenced from S in-plane and Cr $d$ e$_g$ orbitals to out-of-plane S orbitals when forming the monolayer. These out-of-plane orbitals hybridize and develop pipelines connecting the S atoms in up- and down-layers, respectively, which builds an electron network for itinerant electrons. A super-exchange coupling mechanism is thus suggested between two adjacent Cr atoms.

Here, Cr$^{4+}$ adopts a $p^3sd^2$ hybridization that two $e_g$ orbitals, i.e. $d_{x2-y2}$ and $d_{z^2}$ are involved in forming σ-bonds with S $p$-orbitals. The two remaining $d$-electrons are thus filled into three nearly degenerated $t_{2g}$ orbitals (Fig. 2(e)), leading to a metallic CrS$_2$ monolayer with partially filled $d_{xz}$ and $d_{yz}$ orbitals (Fig. 2(g-h)). We may regard it as a highly *p*-doped semiconductor because of an unfilled bandgap of 0.77 eV sitting 1.1 eV above the Fermi Level ($E_F$). Orbital decompositions were mapped on the bandstructures (Fig. 2(g-h)). All five *d* orbitals are unoccupied and grouped together for the spin-down component, however, they split into two groups for the spin-up component. It confirms the $p^3sd^2$ hybridization picture and shows the sigma bonding (~ 3 eV below $E_F$) and anti-bonding (~ 2eV above $E_F$) states (green and cyan dots). Spins of the two *d*-electrons filling the $t_{2g}$ orbitals, together with back-donated electrons from the Cr-S σ-bonds, are all parallel aligned (Fig. 2(e)) owing to the Hund's rule, which results in a local magnetic moment of 2.8 μ$_B$. Spin-orbit coupling does not qualitatively change the bandstructure, but further splits the two spin components by 74 meV (Supplementary Fig. S3 [58]). The super-exchange mechanism might be suppressed by transferring additional electrons into the $t_{2g}$ orbitals. In that case, the mixed valence states of $Cr^{4+}$ and $Cr^{3+}$ promotes the double-exchange mechanism favoring a FM long-range ordering (Fig. 2(f)).

## C. Manipulation of magnetic ordering by additional CrS$_2$ layers.

Interlayer coupling of multilayers could redistribute charge density at the interlayer region [32, 69]. We thus examined a CrS$_2$ bilayer shown in Fig. 3(a) where the AA stacking was found 0.34 eV more stable than the AB stacking (Supplementary Fig. S4 & Table S2 [58]). Spin-exchange coupling parameters (marked in Fig 3a) were extracted by calculating the total energy differences of eight magnetic configurations (Fig.3b-i) based on the Heisenberg Model. The energy contributed by magnetic interaction in these magnetic orders in a unit-cell are expressed as:

$$E_b = \frac{N^2}{4} \times \frac{1}{2}(6J_1 + 6J_2 + 6J_3 + J_4 + 6J_5 + 6J_6)$$

$$E_c = \frac{N^2}{4} \times \frac{1}{2}(-2J_1 + 6J_2 - 2J_3 + J_4 - 2J_5 - 2J_6)$$

$$E_d = \frac{N^2}{4} \times \frac{1}{2}(2J_1 - 2J_2 - 2J_3 + J_4 + 2J_5 - 2J_6)$$

$$E_e = \frac{N^2}{4} \times \frac{1}{2}(-J_1 + 2J_2 - J_3 + J_4 - J_5 + J_6)$$

$$E_f = \frac{N^2}{4} \times \frac{1}{2}(6J_1 + 6J_2 + 6J_3 - J_4 - 6J_5 - 6J_6)$$

$$E_g = \frac{N^2}{4} \times \frac{1}{2}(-2J_1 + 6J_2 - 2J_3 - J_4 + 2J_5 + 2J_6)$$

$$E_h = \frac{N^2}{4} \times \frac{1}{2}(2J_1 - 2J_2 - 2J_3 - J_4 - 2J_5 + 2J_6)$$

$$E_i = \frac{N^2}{4} \times \frac{1}{2}(-J_1 + 2J_2 - J_3 - J_4 + J_5 - J_6),$$

where *N* represents the unpaired spins on each Cr atom, which is treated as 2 in our exchange parameter calculations.

Here, the nearest interlayer parameter $J_4$ was derived -10.8 meV, considerably favoring FM, while $J_1$ and $J_2$ reduce to 16.3 meV and -19.4 meV, respectively. Nevertheless, another three parameters are negligible (<1 meV). It is exceptional that $J_4$ is over an order of magnitude larger than those of $Cr_2Ge_2Te_6$ [29], $RuCl_3$ [44] and $MnI_2$[45], which is also robust under different U values, i.e. 2.0, 2.8 and 4.0 eV (see Supplementary Fig. S1b [58]). Unlike the monolayer case, the bilayer exceptionally undergoes a transition from the intra-layer sAFM order in the monolayer to a both intra- and inter-layer FM order, with at least 8 meV/Cr energy-gain, in bilayer and thicker layers (Supplementary Table S2 and S3 [58]). This FM-FM configuration offers the shortest interlayer S-S distance of 3.18 Å, 0.04 Å smaller than the second shortest distance among all configurations. After this transition, the three-fold symmetry restores and the Cr-S bond lengths are all 2.38 Å.

The bilayer contains four S sub-layers and two Cr layers. Figure 3(j) shows an interlayer DCD with an isosurface value of 0.0005 $e$/Bohr$^3$. This value is over twice those used in plotting $PtS_2$ and $MoS_2$ in Ref. [32], suggesting a rather strong electronic hybridization between the second and third S sub-layers (balls in orange). Apparent charge reduction (green) was found near these two sub-layers and charge accumulation (light-rose) occurs around the middle region of them (marked by the blue rectangle). It shows that a considerable portion of electrons are shared by those two S sub-layers, weakening the existed intra-layer Cr-S bonding. As a result, a noticeable charge transfer from $e_g$ to $t_{2g}$ orbitals is observable [Fig. 3(k)], resulting in a mixture of $Cr^{4+}$ and $Cr^{3+}$ in the bilayer, as schematically shown in Fig. 2(f). This state, together with delocalized S $p$ orbitals and their resulting strong interlayer S-S hopping, favors the double-exchange mechanism, giving rise to a large $J_2$ and $J_4$ for ferromagnetic coupling.

The magnetic anisotropy energies (MAE) of 1L to 4L were considered by comparing the total energies with the magnetic moments parallel to five different vectors ***a***, ***b***, ***c***, ***P*** *and* ***I*** (Supplementary Table S4 and Fig. S5 [58]). The found sAFM-FM transition is still valid in these non-collinear-spin calculations. The easy magnetization axis oscillates between out- (odd number of layers) and in- (even) plane directions, although the differences among vectors **b**, **I** and **P** are even smaller than 0.01 meV and are thus indistinguishable in 2L. Nevertheless, either **I** or **P** direction could be regarded as a tilted in-plane direction. Such oscillation is, most likely, resulted from a confined quantum well state formed between top and bottom layers [70, 71] as reported in Co few-layers[72]. The bandstructure of the bilayer shows a metallic feature, but the $E_F$ is slightly shifted by ~0.1 eV (Fig. 3(l)) compared to the monolayer case. In addition, the interlayer force constant of the bilayer was derived $52.6 \times 10^{18}$ N/m$^3$ and $33.7 \times 10^{18}$ N/m$^3$ for the breathing and shear modes, similar to the values of the breathing mode of $PtS_2$[32] and the shear mode of $MoS_2$[73], respectively. In light of these results, $CrS_2$ appears strong electronic and magnetic but moderate mechanical interlayer-couplings, which imply the feasibility of tuning magnetism by electric controlled charge manipulation.

### D. Manipulation of in-plane magnetic ordering by charge doping.

Direct charge doping using an ionic potential method [66] efficiently tunes magnetism in few-layer $CrS_2$ that $n$-doping leads to an explicit $e_g$ to $t_{2g}$ charge transfer, as illustrated in a doping DCD of a monolayer [Fig. 4(a)]. A $p$-doping case shows an opposite pattern (Fig. 5(b)), implying that a FM-to-sAFM transition in the bilayer is inducible by $p$-doping. Figure 4(c) plots the relative energies of sAFM and FM as a function of doping level for both mono- and bi-layers. It shows *p*-doping is always prone to make the sAFM ordering more favored while *n*-doping plays an opposite role. In the monolayer, an *n*-doping level of 0.13 *e*/S leads to the both configurations energetically degenerate. The largest energy differences of 25 meV and 54 meV achieve at 0.35 and 0.25 *e*/S for mono- and bi-layers, respectively. After these doping levels, the difference decreases and nearly vanishes at 0.5 to 0.6 *e*/S. These results, again, support the mixed valence picture for double exchange. Ideally, the 0.25 *e*/S doping corresponds to the half $Cr^{3+}$ and half $Cr^{4+}$ case, which maximizes the double exchange strength. For the level at 0.5 *e*/S, however, all $Cr^{4+}$ were reduced to $Cr^{3+}$, which suppresses the double-exchange coupling. Charge doping also shifts the $E_F$ that changes the monolayer from a heavily p-doped sAFM semiconductor, to a half-metal, then an FM semiconductor with a bandgap of 2.24 eV and eventually to a n-type semiconductor (FIG. 4(d-g)). The bilayer shows a similar trend as shown in Supplementary Fig. S6 [58]. Another practical charge doping route lies in elemental substitution of S by Cl, as inspired by the recently synthesized Monolayer Janus [46, 47]. Figure 4(h) shows a CrSCl Janus monolayer where all S atoms in a sublayer are entirely replaced by Cl atoms. It is an intrinsic FM semiconductor with a bandgap of 2.11 eV (Fig. 4(i)). The FM state is energetically favored by at least 85 meV/Cr.

Since charge doping could tune both the mono- and the bi-layers from heavily p-doped to n-doped semiconductors, we estimated phonon-limited mobilities for both hole and electron in $CrS_2$ mono- and bi-layers (Table I). The hole mobilities of the two in-plane directions in both 1L and 2L are in hundreds of $cm^2/V \cdot s$, comparable with those of BP. The electron mobilities are generally larger than hole mobilities, especially the electron mobility of 2L $CrS_2$ of over ten-thousand $cm^2/V \cdot s$ in the both in-plane directions. The conduction band of $CrS_2$ layers is comprised of Cr $d_{z^2}$ orbitals, which is less affected by the in-plane structural deformation. This feature, therefore, leads to a rather small deformation potential for electron, giving rise to high electron mobilities.

## E. Proposed devices.

In light of these, it opens diverse possibility to manipulate CrS$_2$ few-layers transiting between sAFM and FM states, which could be employed in magnetic data recording and information processing with the manipulation by, e.g., sliding of the second layer (Fig. 5(a) and 5(b)) and electric field (Fig. 5(c) and 5(d)). As illustrated in Fig. 5(a) and 5(b), the stacking of an additional CrS$_2$ layer could, most likely, maintain the FM ordering locally. The sAFM and FM states, distinguishable by with or without net magnetic moment, could be employed for data recording or processing, i.e. sAFM standing for bit ``0'' and FM for ``1''. In Fig. 5(a), we assume three bits that reads ``100''. They, however, changes to ``010'' if the second layer was moved from the left to the middle position by, e.g., an atomic force microscopy tip. Another route for charge doping lies in electric field applied to shift chemical potentials of samples. Fig. 5(c) shows a schematic diagram of a magnetic field effect transistor where the gate voltage could tune the CrS$_2$ few-layers switching between sAFM and FM states. Spin polarized current thus flows differently in this device through two spin components if it maintains the FM state. In line with this idea, an array could be built as illustrated in Fig. 5(d). This array could be used as data recording that each gate controls each bit switching between the FM (1) and sAFM (0) states.

## F. CONCLUSION

In summary, we show intra-layer magnetism is tunable by interlayer engineering, direct charge doping and elemental substitution, as illustrated in CrS$_2$ few-layers. In particular, an in-plane sAFM-to-FM transition undergoes when the thickness increases from mono- to multi-layers where the interlayer coupling strongly favors FM ($J_4$ = -10.8 meV). The coexistence of Cr$^{4+}$ and Cr$^{3+}$ induced by interlayer charge sharing favors the FM in-plane coupling via a double-exchange coupling mechanism, which is confirmed by our calculations of the doped CrS$_2$ mono- and bi-layers. The CrSCl Janus monolayer, an example of substitutional charge doping, was predicted to be an intrinsic FM semiconductor. Our results manifest that interlayer coupling could, exceptionally, tune magnetic properties, in addition to its tunability of electronic[4, 30-32, 34], mechanical[4, 30], vibrational[14] and optical[13, 33] properties. It opens diverse possibility to manipulate CrS$_2$ few-layers transferring between sAFM and FM states, which could be employed in magnetic data recording and information processing, as proposed in prototypical devices (Fig. 5). Here, either layer stacking or charge doping changes intralayer magnetism, suggesting the magnetic interactions in both intra- and inter-layer directions are still coupled even in the bi-layer limit of CrS$_2$, a strong electronic coupling material. We would expect that the magnetic interactions may be decoupled for both the directions in more weakly interacting bilayers. All these results allow us to manipulate magnetic long-range ordering through various routes, e.g. external electric gating, in two-dimensional materials, which shall boost magnetic applications of 2D materials in nanoelectronics, spintronics and optoelectronics.

# ACKNOWLEDGEMENTS

Project supported by the National Natural Science Foundation of China (Gant Nos. 11274380, 91433103, 11622437, and 61674171), the Fundamental Research Funds for the Central Universities, China and the Research Funds of Renmin University of China (Grant No. 16XNLQ01). C.W was supported by the Outstanding Innovative Talents Cultivation Funded Programs 2017 of Renmin University of China. Calculations were performed at the Physics Lab of High-Performance Computing of Renmin University of China.

# REFERENCES

[1] A.H. Castro Neto, F. Guinea, N.M.R. Peres, K.S. Novoselov and A.K. Geim, The electronic properties of graphene, Rev. Mod. Phys. **81**, 109-162 (2009).
[2] A.K. Geim and K.S. Novoselov, The rise of graphene, Nat Mater **6**, 183-191 (2007).
[3] W. Han, R.K. Kawakami, M. Gmitra and J. Fabian, Graphene spintronics, Nat. Nanotechnol. **9**, 794-807 (2014).
[4] J.-B. Wu, Z.-X. Hu, X. Zhang, W.-P. Han, Y. Lu, W. Shi, X.-F. Qiao, M. Ijiäs, S. Milana, W. Ji, A.C. Ferrari and P.-H. Tan, Interface Coupling in Twisted Multilayer Graphene by Resonant Raman Spectroscopy of Layer Breathing Modes, ACS Nano **9**, 7440-7449 (2015).
[5] M. Chhowalla, H.S. Shin, G. Eda, L.-J. Li, K.P. Loh and H. Zhang, The chemistry of two-dimensional layered transition metal dichalcogenide nanosheets, Nat. Chem, **5**, 263-275 (2013).
[6] K.F. Mak, K.L. McGill, J. Park and P.L. McEuen, The valley Hall effect in MoS2; transistors, Science

**344**, 1489 (2014).

[7] S. Barja, S. Wickenburg, Z.-F. Liu, Y. Zhang, H. Ryu, M.M. Ugeda, Z. Hussain, Z.-X. Shen, S.-K. Mo, E. Wong, M.B. Salmeron, F. Wang, M.F. Crommie, D.F. Ogletree, J.B. Neaton and A. Weber-Bargioni, Charge density wave order in 1D mirror twin boundaries of single-layer $MoSe_2$, Nature Phys. **12**, 751-756 (2016).

[8] J. Lin, S.T. Pantelides and W. Zhou, Vacancy-Induced Formation and Growth of Inversion Domains in Transition-Metal Dichalcogenide Monolayer, ACS Nano **9**, 5189-5197 (2015).

[9] H. Liu, L. Jiao, F. Yang, Y. Cai, X. Wu, W. Ho, C. Gao, J. Jia, N. Wang, H. Fan, W. Yao and M. Xie, Dense Network of One-Dimensional Midgap Metallic Modes in Monolayer MoSe2 and Their Spatial Undulations, Phy. Rev. Lett. **113**, 066105 (2014).

[10] H. Liu, A.T. Neal, Z. Zhu, Z. Luo, X. Xu, D. Tománek and P.D. Ye, Phosphorene: An Unexplored 2D Semiconductor with a High Hole Mobility, ACS Nano **8**, 4033-4041 (2014).

[11] L. Li, Y. Yu, G.J. Ye, Q. Ge, X. Ou, H. Wu, D. Feng, X.H. Chen and Y. Zhang, Black phosphorus field-effect transistors, Nat. Nanotechnol. **9**, 372-377 (2014).

[12] J. Qiao, X. Kong, Z.-X. Hu, F. Yang and W. Ji, High-mobility transport anisotropy and linear dichroism in few-layer black phosphorus, Nat. Commun. **5**, 4475 (2014).

[13] Q. Jia, X. Kong, J. Qiao and W. Ji, Strain- and twist-engineered optical absorption of few-layer black phosphorus, Sci. China Ser. G. **59**, 696811 (2016).

[14] Z.-X. Hu, X. Kong, J. Qiao, B. Normand and W. Ji, Interlayer electronic hybridization leads to exceptional thickness-dependent vibrational properties in few-layer black phosphorus, Nanoscale **8**, 2740-2750 (2016).

[15] N. Mao, J. Tang, L. Xie, J. Wu, B. Han, J. Lin, S. Deng, W. Ji, H. Xu, K. Liu, L. Tong and J. Zhang, Optical Anisotropy of Black Phosphorus in the Visible Regime, J. Am. Chem. Soc. **138**, 300-305 (2016).

[16] O.V. Yazyev and L. Helm, Defect-induced magnetism in graphene, Phys. Rev. B **75**, 125408 (2007).

[17] H. González-Herrero, J.M. Gómez-Rodríguez, P. Mallet, M. Moaied, J.J. Palacios, C. Salgado, M.M. Ugeda, J.-Y. Veuillen, F. Yndurain and I. Brihuega, Atomic-scale control of graphene magnetism by using hydrogen atoms, Science **352**, 437 (2016).

[18] K.M. McCreary, A.G. Swartz, W. Han, J. Fabian and R.K. Kawakami, Magnetic Moment Formation in Graphene Detected by Scattering of Pure Spin Currents, Phy. Rev. Lett. **109**, 186604 (2012).

[19] B. Uchoa, V.N. Kotov, N.M.R. Peres and A.H. Castro Neto, Localized Magnetic States in Graphene, Phy. Rev. Lett. **101**, 026805 (2008).

[20] J. Zhou, Q. Wang, Q. Sun, X.S. Chen, Y. Kawazoe and P. Jena, Ferromagnetism in Semihydrogenated Graphene Sheet, Nano Lett. **9**, 3867-3870 (2009).

[21] C. Zhao, C. Jin, J. Wu and W. Ji, Magnetism in molybdenum disulphide monolayer with sulfur substituted by 3d transition metals, J. Appl. Phys. **120**, 144305 (2016).

[22] R.R. Nair, M. Sepioni, I.L. Tsai, O. Lehtinen, J. Keinonen, A.V. Krasheninnikov, T. Thomson, A.K. Geim and I.V. Grigorieva, Spin-half paramagnetism in graphene induced by point defects, Nature Phys. **8**, 199-202 (2012).

[23] J. Cervenka, M.I. Katsnelson and C.F.J. Flipse, Room-temperature ferromagnetism in graphite driven by two-dimensional networks of point defects, Nature Phys. **5**, 840-844 (2009).

[24] J. Jung, T. Pereg-Barnea and A.H. MacDonald, Theory of Interedge Superexchange in Zigzag Edge Magnetism, Phy. Rev. Lett. **102**, 227205 (2009).

[25] O. Hod, V. Barone, J.E. Peralta and G.E. Scuseria, Enhanced Half-Metallicity in Edge-Oxidized Zigzag Graphene Nanoribbons, Nano Lett. **7**, 2295-2299 (2007).

[26] N.D. Mermin and H. Wagner, Absence of Ferromagnetism or Antiferromagnetism in One- or Two-Dimensional Isotropic Heisenberg Models, Phy. Rev. Lett. **17**, 1133-1136 (1966).
[27] B. Huang, G. Clark, E. Navarro-Moratalla, D.R. Klein, R. Cheng, K.L. Seyler, D. Zhong, E. Schmidgall, M.A. McGuire, D.H. Cobden, W. Yao, D. Xiao, P. Jarillo-Herrero and X. Xu, Layer-dependent ferromagnetism in a van der Waals crystal down to the monolayer limit, Nature **546**, 270-273 (2017).
[28] M.A. McGuire, H. Dixit, V.R. Cooper and B.C. Sales, Coupling of Crystal Structure and Magnetism in the Layered, Ferromagnetic Insulator $CrI_3$, Chemistry of Materials **27**, 612-620 (2015).
[29] C. Gong, L. Li, Z. Li, H. Ji, A. Stern, Y. Xia, T. Cao, W. Bao, C. Wang, Y. Wang, Z.Q. Qiu, R.J. Cava, S.G. Louie, J. Xia and X. Zhang, Discovery of intrinsic ferromagnetism in two-dimensional van der Waals crystals, Nature **546**, 265-269 (2017).
[30] K. Liu, L. Zhang, T. Cao, C. Jin, D. Qiu, Q. Zhou, A. Zettl, P. Yang, S.G. Louie and F. Wang, Evolution of interlayer coupling in twisted molybdenum disulfide bilayers, Nat. Commun. **5**, 4966 (2014).
[31] C. Zhang, C.-P. Chuu, X. Ren, M.-Y. Li, L.-J. Li, C. Jin, M.-Y. Chou and C.-K. Shih, Interlayer couplings, Moiré patterns, and 2D electronic superlattices in $MoS_2/WSe_2$ hetero-bilayers, Sci. Adv. **3**, 1601459 (2017).
[32] Y. Zhao, J. Qiao, P. Yu, Z. Hu, Z. Lin, S.P. Lau, Z. Liu, W. Ji and Y. Chai, Extraordinarily Strong Interlayer Interaction in 2D Layered $PtS_2$, Adv Mater **28**, 2399-2407 (2016).
[33] H.-P. Komsa and A.V. Krasheninnikov, Electronic structures and optical properties of realistic transition metal dichalcogenide heterostructures from first principles, Phys. Rev. B **88**, 085318 (2013).
[34] P.-C. Yeh, W. Jin, N. Zaki, J. Kunstmann, D. Chenet, G. Arefe, J.T. Sadowski, J.I. Dadap, P. Sutter, P. Hone and R.M. Osgood, Direct Measurement of the Tunable Electronic Structure of Bilayer $MoS_2$ by Interlayer Twist, Nano Lett. **16**, 953-959 (2016).
[35] H. Kasai, K. Tolborg, M. Sist, J. Zhang, V.R. Hathwar, M.Ø. Filsø, S. Cenedese, K. Sugimoto, J. Overgaard, E. Nishibori and B.B. Iversen, X-ray electron density investigation of chemical bonding in van der Waals materials, Nature Materials **17**, 249-252 (2018).
[36] J. Qiao, Y. Pan, F. Yang, C. Wang, Y. Chai and W. Ji, Few-layer Tellurium: one-dimensional-like layered elementary semiconductor with striking physical properties, Science Bulletin **63**, 159-168 (2018).
[37] Y. Zhao, J. Qiao, Z. Yu, P. Yu, K. Xu, P. Lau Shu, W. Zhou, Z. Liu, X. Wang, W. Ji and Y. Chai, High-Electron-Mobility and Air-Stable 2D Layered PtSe2 FETs, Advanced Materials **29**, 1604230 (2016).
[38] X. Qiu and W. Ji, Illuminating interlayer interactions, Nature Materials **17**, 211-213 (2018).
[39] H.-B. Li, N. Lu, Q. Zhang, Y. Wang, D. Feng, T. Chen, S. Yang, Z. Duan, Z. Li, Y. Shi, W. Wang, W.-H. Wang, K. Jin, H. Liu, J. Ma, L. Gu, C. Nan and P. Yu, Electric-field control of ferromagnetism through oxygen ion gating, Nat. Commun. **8**, 2156 (2017).
[40] H.L. Zhuang, M.D. Johannes, M.N. Blonsky and R.G. Hennig, Computational prediction and characterization of single-layer $CrS_2$, Appl. Phys. Lett. **104**, 022116 (2014).
[41] J. Zhang, H. Zheng, R. Han, X. Du and Y. Yan, Tuning magnetic properties of CrS2 monolayer by doping transition metal and alkaline-earth atoms, Journal of Alloys and Compounds **647**, 75-81 (2015).
[42] C. Ataca, H. Şahin and S. Ciraci, Stable, Single-Layer $MX_2$ Transition-Metal Oxides and Dichalcogenides in a Honeycomb-Like Structure, The Journal of Physical Chemistry C **116**, 8983-8999 (2012).
[43] S. Lebègue, T. Björkman, M. Klintenberg, R.M. Nieminen and O. Eriksson, Two-Dimensional Materials from Data Filtering and Ab Initio Calculations, Physical Review X **3**, 031002 (2013).
[44] S. Sinn, C.H. Kim, B.H. Kim, K.D. Lee, C.J. Won, J.S. Oh, M. Han, Y.J. Chang, N. Hur, H. Sato, B.-G.

Park, C. Kim, H.-D. Kim and T.W. Noh, Electronic Structure of the Kitaev Material α-RuCl3 Probed by Photoemission and Inverse Photoemission Spectroscopies, **6**, 39544 (2016).

[45] X. Wu, Y. Cai, Q. Xie, H. Weng, H. Fan and J. Hu, Magnetic ordering and multiferroicity in $MnI_2$, Phys. Rev. B **86**, 134413 (2012).

[46] A.-Y. Lu, H. Zhu, J. Xiao, C.-P. Chuu, Y. Han, M.-H. Chiu, C.-C. Cheng, C.-W. Yang, K.-H. Wei, Y. Yang, Y. Wang, D. Sokaras, D. Nordlund, P. Yang, D.A. Muller, M.-Y. Chou, X. Zhang and L.-J. Li, Janus monolayers of transition metal dichalcogenides, Nat. Nanotechnol. **12**, 744–749 (2017).

[47] J. Zhang, S. Jia, I. Kholmanov, L. Dong, D. Er, W. Chen, H. Guo, Z. Jin, V.B. Shenoy, L. Shi and J. Lou, Janus Monolayer Transition-Metal Dichalcogenides, ACS Nano **11**, 8192-8198 (2017).

[48] P.E. Blöchl, Projector augmented-wave method, Phys. Rev. B **50**, 17953-17979 (1994).

[49] G. Kresse and J. Furthmüller, Efficient iterative schemes for ab initio total-energy calculations using a plane-wave basis set, Phys. Rev. B **54**, 11169-11186 (1996).

[50] J. Klimeš, D.R. Bowler and A. Michaelides, Van der Waals density functionals applied to solids, Phys. Rev. B **83**, 195131 (2011).

[51] I.A. Vladimir, F. Aryasetiawan and A.I. Lichtenstein, First-principles calculations of the electronic structure and spectra of strongly correlated systems: the LDA + U method, Journal of Physics: Condensed Matter **9**, 767 (1997).

[52] I.V. Solovyev, P.H. Dederichs and V.I. Anisimov, Corrected atomic limit in the local-density approximation and the electronic structure of d impurities in Rb, Phys. Rev. B **50**, 16861-16871 (1994).

[53] M. Cococcioni and S. de Gironcoli, Linear response approach to the calculation of the effective interaction parameters in the LDA+U method, Phys. Rev. B **71**, 035105 (2005).

[54] B.-T. Wang, W. Yin, W.-D. Li and F. Wang, First-principles study of pressure-induced phase transition and electronic property of $PbCrO_3$, J. Appl. Phys. **111**, 013503 (2012).

[55] M. Aykol and C. Wolverton, Local environment dependent GGA+U method for accurate thermochemistry of transition metal compounds, Phys. Rev. B **90**, 115105 (2014).

[56] A. Jain, G. Hautier, S.P. Ong, C.J. Moore, C.C. Fischer, K.A. Persson and G. Ceder, Formation enthalpies by mixing GGA and GGA + U calculations, Phys. Rev. B **84**, 045115 (2011).

[57] M. Korotin, V. Anisimov, D. Khomskii and G. Sawatzky, $CrO_2$: a self-doped double exchange ferromagnet, Phy. Rev. Lett. **80**, 4305 (1998).

[58] See Supplemental Material at [URL will be inserted by publisher] for more information about the theoretical methods and data analysis.

[59] J. Sugiyama, H. Nozaki, I. Umegaki, T. Uyama, K. Miwa, J.H. Brewer, S. Kobayashi, C. Michioka, H. Ueda and K. Yoshimura, Static magnetic order on the metallic triangular lattice in $CrSe_2$ detected by μSR+, Phys. Rev. B **94**, 014408 (2016).

[60] J. Rouxel, A. Meerschaut and G.A. Wiegers, Chalcogenide misfit layer compounds, Journal of Alloys and Compounds **229**, 144-157 (1995).

[61] C.M. Fang, C.F.v. Bruggen, R.A.d. Groot, G.A. Wiegers and C. Haas, The electronic structure of the metastable layer compound, Journal of Physics: Condensed Matter **9**, 10173 (1997).

[62] S. Takagi, A. Toriumi, M. Iwase and H. Tango, On the universality of inversion layer mobility in Si MOSFET's: Part II-effects of surface orientation, IEEE Trans. Electron Devices **41**, 2363-2368 (1994).

[63] G. Fiori and G. Iannaccone, Multiscale Modeling for Graphene-Based Nanoscale Transistors, Proc. IEEE **101**, 1653-1669 (2013).

[64] P. Kang, V. Michaud-Rioux, X.H. Kong, G.H. Yu and H. Guo, Calculated carrier mobility of h-BN/ γ - InSe/h-BN van der Waals heterostructures, 2D Materials **4**, 045014 (2017).

[65] P. Kang, W.-T. Zhang, V. Michaud-Rioux, X.-H. Kong, C. Hu, G.-H. Yu and H. Guo, Moiré impurities in twisted bilayer black phosphorus: Effects on the carrier mobility, Phys. Rev. B **96**, 195406 (2017).

[66] W. Ji, Z.-Y. Lu and H. Gao, Electron Core-Hole Interaction and Its Induced Ionic Structural Relaxation in Molecular Systems under X-Ray Irradiation, Phy. Rev. Lett. **97**, 246101 (2006).

[67] J.L. Verble, T.J. Wietling and P.R. Reed, Rigid-layer lattice vibrations and van der waals bonding in hexagonal $MoS_2$, Solid State Commun. **11**, 941-944 (1972).

[68] S. Baroni, S. de Gironcoli, A. Dal Corso and P. Giannozzi, Phonons and related crystal properties from density-functional perturbation theory, Rev. Mod. Phys. **73**, 515-562 (2001).

[69] H. Fang, C. Battaglia, C. Carraro, S. Nemsak, B. Ozdol, J.S. Kang, H.A. Bechtel, S.B. Desai, F. Kronast, A.A. Unal, G. Conti, C. Conlon, G.K. Palsson, M.C. Martin, A.M. Minor, C.S. Fadley, E. Yablonovitch, R. Maboudian and A. Javey, Strong interlayer coupling in van der Waals heterostructures built from single-layer chalcogenides, Proceedings of the National Academy of Sciences **111**, 6198-6202 (2014).

[70] M. Przybylski, M. Dąbrowski, U. Bauer, M. Cinal and J. Kirschner, Oscillatory magnetic anisotropy due to quantum well states in thin ferromagnetic films, J. Appl. Phys. **111**, 07C102 (2012).

[71] J. Li, G. Chen, Y.Z. Wu, E. Rotenberg and M. Przybylski, Quantum Well States and Oscillatory Magnetic Anisotropy in Ultrathin Fe Films, IEEE Trans. Magn. **47**, 1603-1609 (2011).

[72] H. Yang, A.D. Vu, A. Hallal, N. Rougemaille, J. Coraux, G. Chen, A.K. Schmid and M. Chshiev, Anatomy and Giant Enhancement of the Perpendicular Magnetic Anisotropy of Cobalt–Graphene Heterostructures, Nano Lett. **16**, 145-151 (2016).

[73] Y. Zhao, X. Luo, H. Li, J. Zhang, P.T. Araujo, C.K. Gan, J. Wu, H. Zhang, S.Y. Quek, M.S. Dresselhaus and Q. Xiong, Interlayer Breathing and Shear Modes in Few-Trilayer $MoS_2$ and $WSe_2$, Nano Lett. **13**, 1007-1015 (2013).

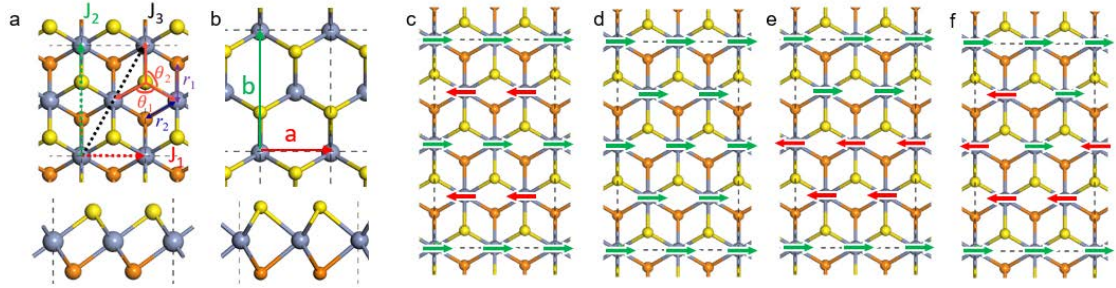

FIG. 1. Schematic models of monolayer $CrS_2$ and four different magnetic orders. (a-b) Top and side view of single layer $CrS_2$ in 1T (**a**) and 2H (**b**) phase. The slate-blue, yellow and orange balls represent Cr, top S and bottom S atoms, respectively. The intralayer spin-exchange parameters $J_1$, $J_2$, $J_3$ between Cr sites are represented with dashed arrows. The red, green, blue and purple arrows correspond to the lattice constant $a$ and $b$, Cr-S bond $r_1$ and $r_2$. The two red arcs represent the Cr-S-Cr angles $\theta_1$ and $\theta_2$, respectively. (c-f) Schematic representation of four intralayer magnetic orders used for calculation of exchange parameters. The green and red arrow represent the magnetic moment up and down on Cr atoms, respectively.

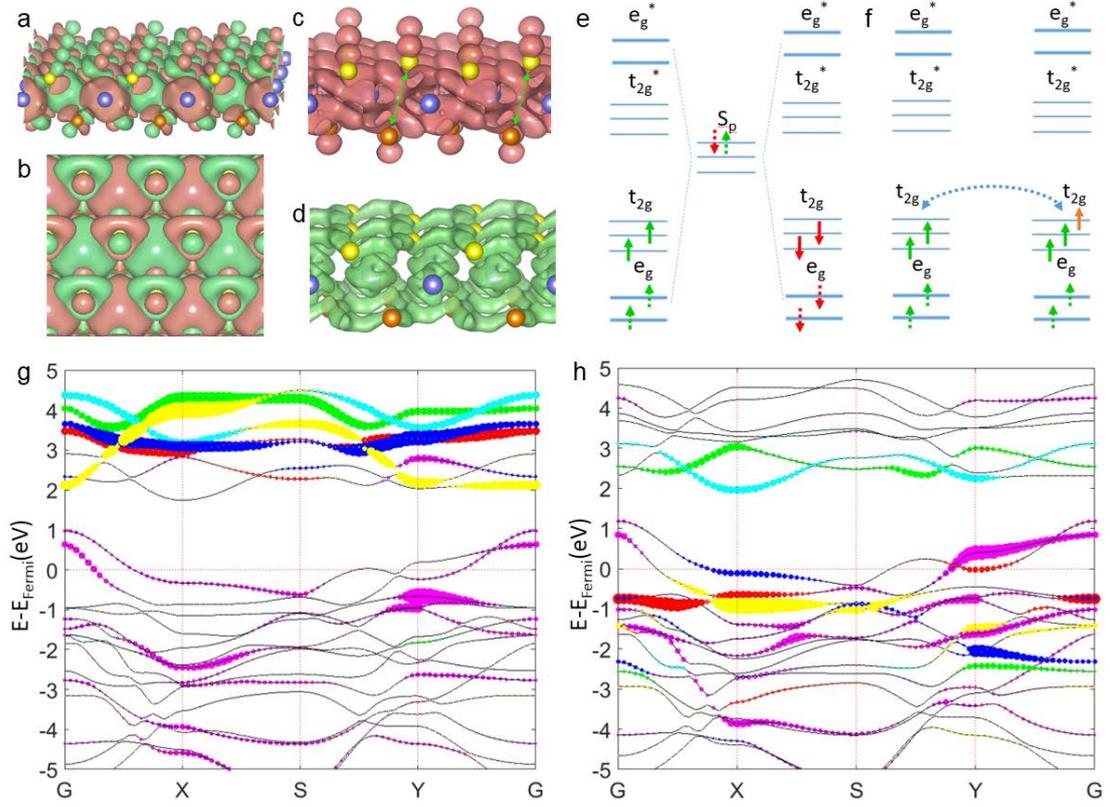

FIG. 2. Electronic structure of sAFM monolayer $CrS_2$. (a-b) Spin charge density map of $CrS_2$ in sAFM order, side and top view. The red and green isosurface correspond to charge with different spin polarize direction up and down. (c-d) Atomic differential charge density of monolayer $CrS_2$. **c** and **d** correspond to the charge accumulation and reduction after Cr and S atoms bonding together respectively. (e-f) Schematics of super-exchange and double-exchange mechanism for spin-exchange between localized *d* electrons of Cr and itinerant electrons comprised of S *p* state and Cr *d* states. The up and down arrows represent the electron with different spin component and the dashed arrows correspond to bonding electrons from Cr and S atoms. The orange arrow in **e** indicates the transferred electrons from layer stacking or doping. (g-h) Electronic band structure of the sAFM monolayer $CrS_2$ with different spin component down (e) and up (f). The Fermi energy is zero. The Cr-d and S-p orbitals are mapped with different colors: Cr-$d_{xy}$, green; Cr-$d_{yz}$, red; Cr-$d_{xz}$, blue; Cr-$d_{z^2}$, cyan; Cr-$d_{x^2-y^2}$, yellow; S-*p*, magenta.

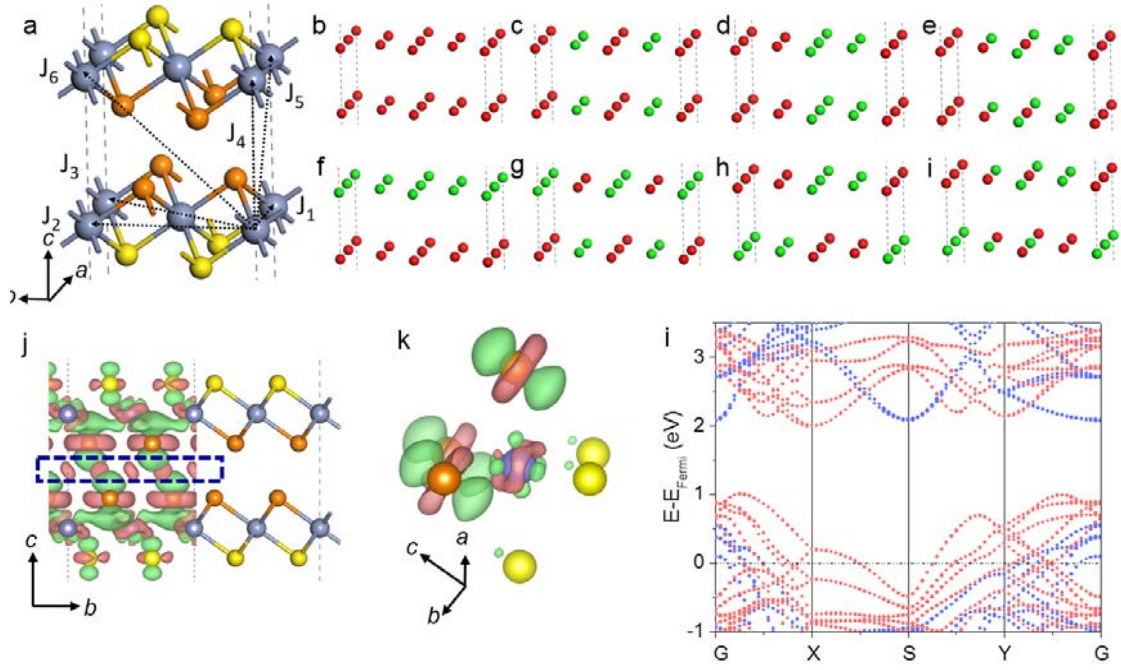

FIG. 3. Bilayer CrS$_2$. (a) Perspective view of AA-stacked bilayer CrS$_2$. Exchange parameters are marked with dashed arrows connected between Cr atoms. (b-i) Schematic representation of eight magnetic orders used for the calculation of exchange parameters in AA-stacked bilayer CrS$_2$. The green and red balls represent the magnetic moment up and down on Cr atoms, respectively. (j) Differential charger density (DCD) of bilayer CrS$_2$ with an isosurface value of 0.0005 $e$/Bohr$^3$. Light-rose and green isosurface indicate charge accumulation and depletion after layer stacking, respectively. (k) A zoomed-in plot around Cr atom of DCD of bilayer CrS$_2$ in (j). (l) Electronic band structure of the FM-FM bilayer CrS$_2$ with different spin component. Red: spin up; blue: spin down.

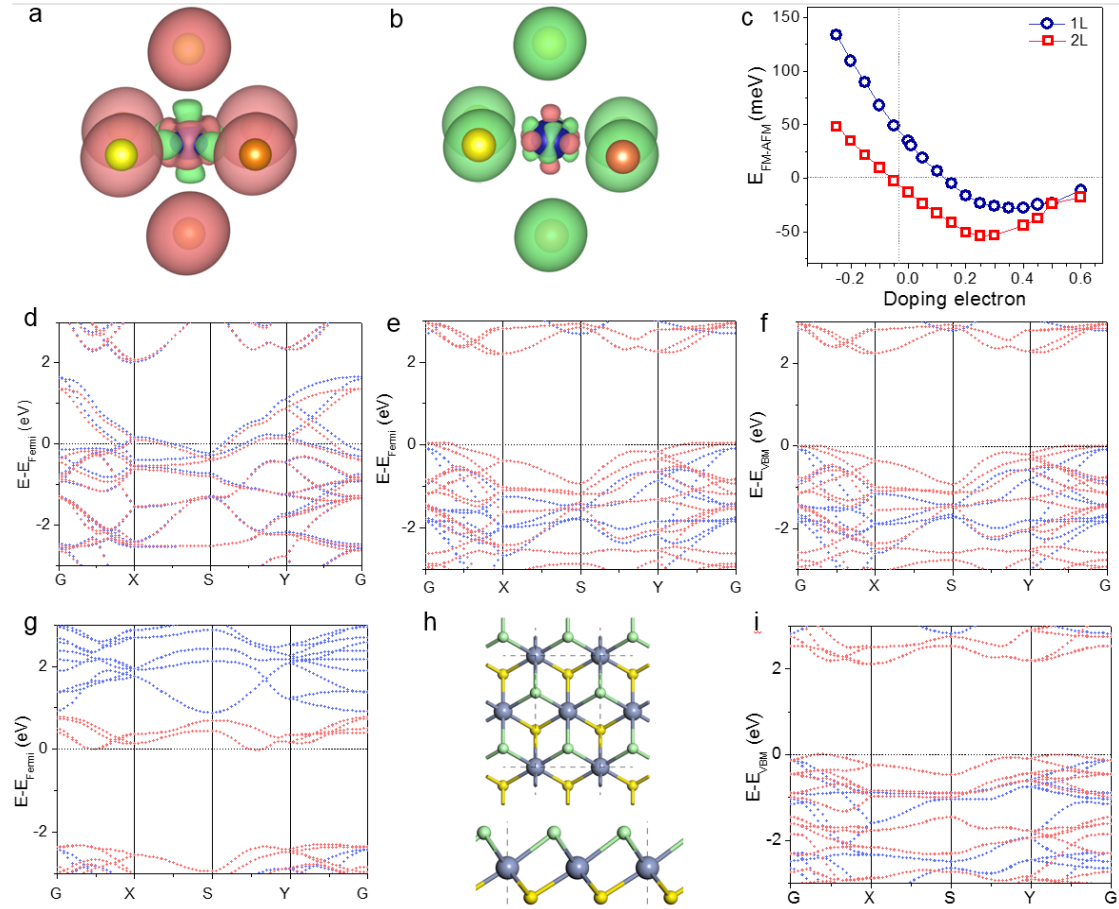

FIG. 4. Manipulation of in-plane magnetic ordering and electronic structure by electron and hole doping. (a-b) Differential charge density of electron/hole doped monolayer $CrS_2$, respectively. (c) Energy difference between FM and sAFM intra-plane magnetic orders with different doping concentration. The blue and red symbols correspond to relative energy in mono- and bi-layer respectively. Doping negative member of electrons represents to doping corresponding amount of holes. (d-g) Band structures of FM monolayer $CrS2$ with different hole/electron doping concentration 0.20 h/S (d), 0.45 e/S (e), 0.50 e/S (f) and 0.60 e/S (g). (h) Top and side views of CrSCl Janus monolayer in the 1T phase. The light-green balls represent Cl atoms. (i) Electronic band structure of the CrSCl monolayer in the ferromagnetic ordering, showing an intrinsic FM semiconductor with an energy gap of 2.11 eV.

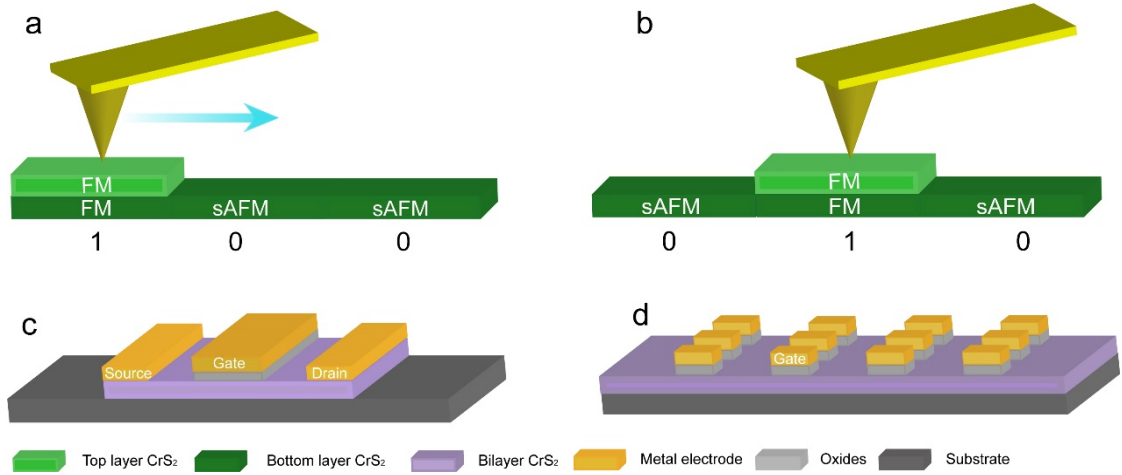

FIG. 5. Schematic drawing of a spin-logic application of few-layer $CrS_2$. (a-b) Manipulation of mono- or bi-layer $CrS_2$ flakes transiting between sAFM and FM states with sliding the top layer. (c) Spin-dependent transfer FET controlled by injected charge from electric gates. (d) Magnetic logic array with every $CrS_2$ domain controlled by the corresponding electric gate.

Table 1
Phonon-limited electron/hole mobility of CrS$_2$ in 2H/1T phase, compared with MoS$_2$, PtSe$_2$[37] and black phosphorus[12]. Here, $m_x^*$ and $m_y^*$ are carrier effective masses along the x and y directions, respectively; E$_{1x}$ (E$_{1y}$), C$_{x\_2D}$ (C$_{y\_2D}$), and μ$_{x-2D}$ (μ$_{y-2D}$) are the deformation potentials, 2D elastic modulus, and the mobility along the x (y) direction, respectively.

| Carrier type | Number of layer | | m$_x$*/m$_0$ | m$_y$*/m$_0$ | m$_d$ | C$_{x\_2D}$ | C$_{y\_2D}$ | E$_{1x}$ | E$_{1y}$ | μ$_{x-2D}$ | μ$_{y-2D}$ |
|---|---|---|---|---|---|---|---|---|---|---|---|
| | | | | | | (J m$^{-2}$) | | (eV) | | (10$^3$cm$^2$ V$^{-1}$ s$^{-1}$) | |
| e | 1L | 2H | 0.90 | 0.84 | 0.87 | 117.57 | 117.91 | 4.28 | 4.57 | 0.17 | 0.17 |
| | | 1T | 1.15 | 0.72 | 0.91 | 73.43 | 63.84 | 0.71 | 2.23 | 2.31 | 0.53 |
| | | MoS$_2$ | 0.47 | 0.47 | | 129.93 | 135.34 | 5.57 | 5.55 | 0.41 | 0.43 |
| | | PtSe$_2$ | 0.26 | 0.48 | | 66.95 | 66.82 | 2.20 | 0.72 | 3.25 | 16.25 |
| | | BP | 0.17 | 1.12 | 0.44 | 14.5 | 50.8 | 2.72 | 7.11 | 0.56 | 0.04 |
| | 2L | 2H | 0.74 | 0.74 | 0.74 | 234.28 | 234.20 | 4.16 | 4.55 | 0.51 | 0.43 |
| | | 1T | 0.33 | 0.49 | 0.40 | 93.82 | 92.02 | 1.05 | 0.70 | 16.46 | 16.71 |
| | | BP | 0.18 | 1.13 | 0.45 | 28.74 | 97.31 | 5.02 | 7.35 | 0.30 | 0.07 |
| h | 1L | 2H | 0.90 | 0.84 | 0.87 | 117.57 | 117.57 | 2.41 | 2.14 | 0.53 | 0.78 |
| | | 1T | 0.76 | 0.28 | 0.46 | 73.43 | 63.84 | 2.23 | 8.90 | 0.54 | 0.22 |
| | | BP | 0.15 | 6.35 | 0.98 | 14.47 | 50.80 | 2.50 | 0.15 | 0.33 | 7.71 |
| | 2L | 2H | 2.74 | 2.40 | 2.57 | 234.28 | 234.20 | 1.43 | 1.31 | 0.29 | 0.46 |
| | | 1T | 0.76 | 2.23 | 1.30 | 93.82 | 92.02 | 1.71 | 1.40 | 1.15 | 0.19 |
| | | BP | 0.15 | 1.81 | 0.52 | 28.74 | 97.31 | 2.45 | 1.63 | 1.30 | 0.82 |

# Supporting Information

# Layer and doping tunable ferromagnetic order in two-dimensional CrS$_2$ layers

Cong Wang[1], Xieyu Zhou[1], Yuhao Pan[1], Jingsi Qiao[1], Xianghua Kong[1], Chao-Cheng Kaun[2], Wei Ji[1,*]

[1]*Beijing Key Laboratory of Optoelectronic Functional Materials & Micro-Nano Devices, Department of Physics, Renmin University of China, Beijing 100872, P.R. China*

[2]*Research Center for Applied Sciences, Academia Sinica, Taipei 11529, Taiwan, R. China*
Email: wji@ruc.edu.cn

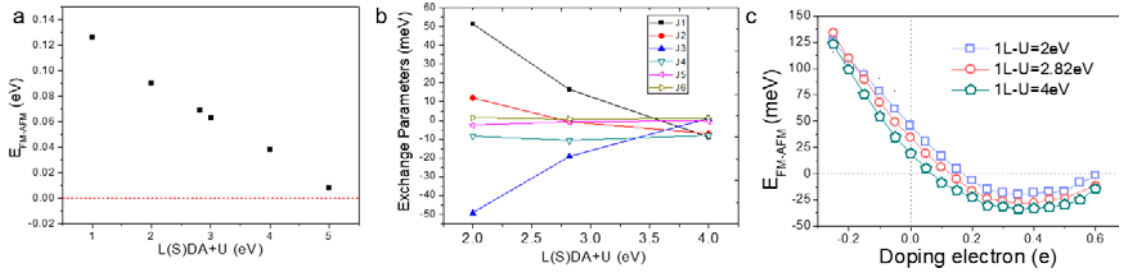

Supplementary Figure S1. Effects of different on-site Coulomb U values. (a) Energy differences between intra-plane FM and sAFM magnetic orders in monolayer $CrS_2$ with respect to different effective U values. (b) Spin exchange parameters in bilayer $CrS_2$ with different U values. (c) Energy difference between the FM and sAFM orders as a function of doping level with different effective U values. These results with U values from 1eV to 5eV suggest the magnetic groundstate of sAFM for monolayer $CrS_2$. We also tested U values of 2 eV and 4 eV for their influence on exchange parameters. The results show that the intra-layer spin exchange parameters $J_1$, $J_2$ and $J_3$ are slightly sensitive to the U values, but the large interlayer spin-exchange parameter $J_4$ does not appreciably change with different U values. The sAFM-FM transition upon charge doping, one of the key results in this work, maintains with U values varying from 2eV to 4eV.

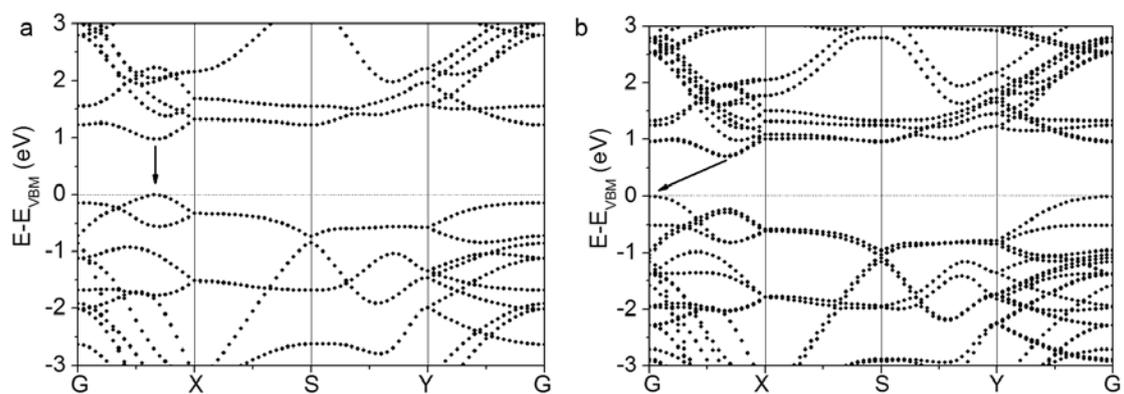

Supplementary Figure S2. Bandstructure of non-magnetic 2H phase mono- (a) and bi-layer $CrS_2$ (b). Similar to $MoS_2$, $CrS_2$ in 2H phase undergoes the transition from direct to indirect semiconductor with the number of layer increased from monolayer to bilayer. The energy gap is 0.97 and 0.71 eV in mono- and bi-layer $CrS_2$ respectively.

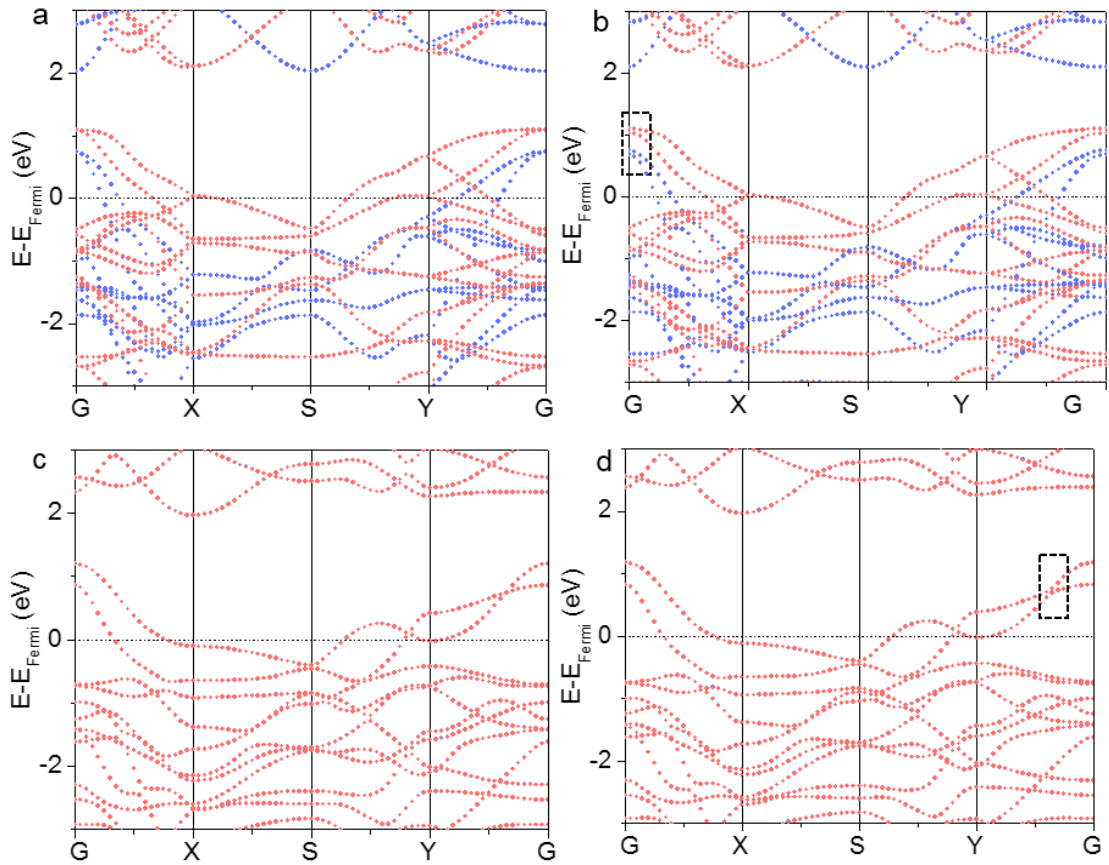

Supplementary Figure S3. Effects of Spin-orbit coupling (SOC) on band structure of FM/sAFM monolayer $CrS_2$. (a)FM without SOC. (b)FM with SOC. (c)sAFM without SOC. (d) sAFM with SOC. For FM (a-b), the band splitting on Γ point induced by SOC is about 78 meV. As for sAFM (c-d), the band splitting induced by SOC is about 74 meV on the point marked by the black dashed rectangle. The influence of SOC on band structure is negligible for few layer $CrS_2$.

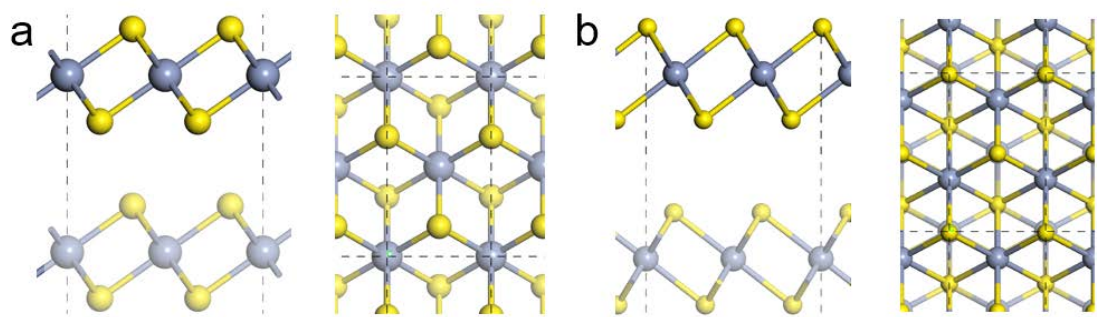

Supplementary Figure S4. Schematic models of 1T bilayer $CrS_2$ with stacking orders AA (a) and AB (b). Stacking orders AA (a) and AB (b) are confirmed to be most stable candidates in few-layer $MoS_2$ and $MoSe_2$ according to previous experiment and calculations. AA stacking was found to be more stable than AB when on site Coulomb U correction is applied, no matter what intra- and inter-layer magnetic orders. Exact data of relative energy can be found in Supplementary Table 2.

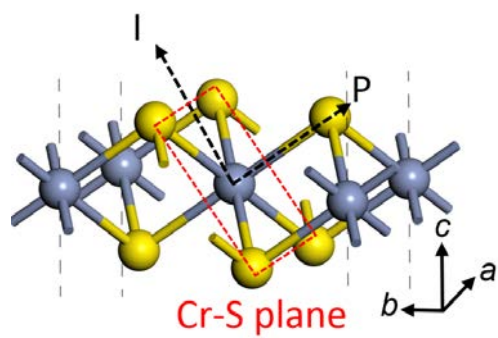

Supplementary Figure S5. Five magnetization axis considered in the calculation of the MAE (magnetic anisotropy energy). Axes *a, b,* and *c* correspond to the directions of lattice vectors. They are perpendicular to each other. A Cr-S plane is marked with a red dashed rectangular in the figure. Axes *I* and *P*, marked with black dashed arrows, represent the directions that are in and perpendicular to the Cr-S plane, respectively.

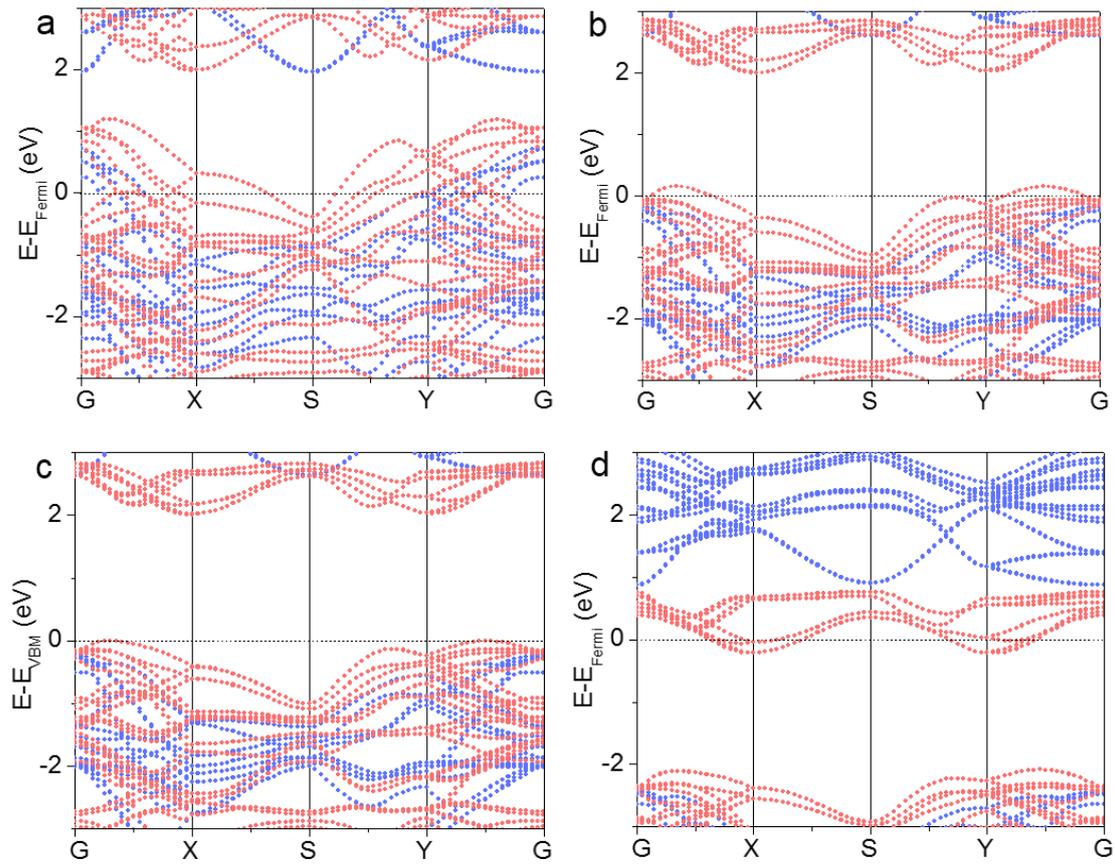

Supplementary Figure S5. Band structure of FM-FM bilayer CrS$_2$ with different electron doping concentrations, including 0.10 *h*/S (a), 0.45 *e*/S (b), 0.5 *e*/S (c) and 0.6 *e*/S (d). Charge doping to the bilayer shows the same trend with monolayer. Electron doping can also tune bilayer CrS$_2$ into semiconductor with a in direct band gap of 2.02 eV (c).

Supplementary Table S1

Relative total energy $\Delta E_0$ (with respect to the most stable configuration calculated with the same effective U values), magnetic moment per Cr atom, lattice constants $a$ and $b$ in monolayer $CrS_2$. A non-magnetic 2H phase was found to be more stable without the consideration of the on-site Coulomb U correction, consistent with previous studies. If the on-site Coulomb interaction is properly treated with DFT+U ($U_{eff}$ = 2.8 eV), a striped anti-ferromagnetic (sAFM) 1T phase was revealed to be the most stable in monolayer $CrS_2$. A Jahn-Teller distortion in sAFM reduces the 3-fold structure symmetry ($a$ = 3.31 Å, $b$ = 5.45 Å), but the symmetry persists in FM ($a$ = 3.29 Å, $b$ = 5.70 Å).

| U (eV) | Configuration | $\Delta E_0$ (eV/f.u.) | Mag. ($\mu_B$) | $a$ (Å) | $b$ (Å) |
|---|---|---|---|---|---|
| 0 | 2H-NM | 0 | 0 | 3.02 | 5.23 |
| 0 | 1T-NM | 0.540 | 0 | 3.03 | 5.25 |
| 0 | 1T-FM | 0.580 | 2.27 | 3.27 | 5.67 |
| 0 | 1T-sAFM | 0.465 | 1.96 | 3.25 | 5.35 |
| 2.8 | 2H-FM | 0.420 | 2.61 | 3.23 | 5.60 |
| 2.8 | 2H-sAFM | 0.325 | 2.38 | 3.17 | 5.46 |
| 2.8 | 1T-FM | 0.034 | 2.82 | 3.29 | 5.70 |
| 2.8 | 1T-sAFM | 0 | 2.72 | 3.31 | 5.45 |

Supplementary Table S2

Relative total energy $\Delta E_0$ (with respect to the most stable configuration AA-FM-FM), the binding energy per formula unit, magnetic moment per Cr, lattice constants *a* and *b*, and the interlayer distance in a CrS$_2$ bilayer. The configurations are named with the form such as "AA-FM-FM", which means a bilayer CrS$_2$ in an AA stacking with the ferromagnetic intra- and inter-layer magnetic order. Both intra- and interlayer ferromagnetic orders are the most stable in the bilayer, accompanied by the restored 3-fold symmetry. The binding energy of configuration AA-FM-FM is -0.34 eV, which is slightly larger than those PtS$_2$ and MoS$_2$.

| Configuration | $\Delta E_0$ (meV/Cr) | E$_b$ (eV/f.u.) | Mag. ($\mu_B$) | *a* (Å) | *b* (Å) | *d* (Å) |
|---|---|---|---|---|---|---|
| AA-FM-FM | 0 | -0.34 | 2.85 | 3.35 | 5.81 | 2.53 |
| AA-FM-AFM | 8 | -0.32 | 2.89 | 3.37 | 5.82 | 2.51 |
| AA-sAFM-FM | 11 | -0.25 | 2.75 | 3.36 | 5.58 | 2.56 |
| AA-sAFM-AFM | 19 | -0.23 | 2.74 | 3.34 | 5.55 | 2.69 |
| AB-FM-FM | 22 | -0.30 | 2.85 | 3.33 | 5.78 | 2.73 |
| AB-FM-AFM | 29 | -0.28 | 2.88 | 3.35 | 5.80 | 2.68 |
| AB-sAFM-FM | 23 | -0.23 | 2.73 | 3.32 | 5.48 | 2.94 |
| AB-sAFM1-AFM | 22 | -0.23 | 2.73 | 3.32 | 5.48 | 2.98 |
| 1L-sAFM | | | 2.72 | 3.31 | 5.45 | |
| PtS$_2$ | | -0.28 | | | | |
| MoS$_2$ | | -0.24 | | | | |

Supplementary Table S3
Relative total energy $\Delta E_0$ (with respect to the most stable configuration AA-FM-FM), interlayer binding energy per formula unit, magnetic moment per Cr, lattice constants $a$ and $b$ and interlayer distance $d$ in a $CrS_2$ trilayer. Here, the FM-FM order is still the most stable configuration.

| Configuration | $\Delta E_0$ (meV/Cr) | $E_b$ (eV/f.u.) | Mag. ($\mu_B$) | $a$ (Å) | $b$ (Å) | $d$ (Å) |
|---|---|---|---|---|---|---|
| AA-FM-FM | 0 | -0.36 | 2.86 | 3.36 | 5.82 | 2.57 |
| AA-FM-AFM | 17 | -0.33 | 2.88 | 3.37 | 5.84 | 2.56 |
| AA-sAFM-FM | 33 | -0.26 | 2.76 | 3.37 | 5.61 | 2.61 |
| AA-sAFM-AFM | 40 | -0.25 | 2.75 | 3.36 | 5.59 | 2.64 |
| 1L-sAFM | | | 2.72 | 3.31 | 5.45 | |
| $PtS_2$ | | -0.28 | | | | |
| $MoS_2$ | | -0.24 | | | | |

Table S4

Magnetic anisotropy energy (MAE) per Cr atom (with respect to the configuration with the most energetically favored magnetization axis) of $CrS_2$ with in-plane FM/sAFM and inter-plane FM from 1L to 4L. Five possible easy magnetization axes were considered for few layer $CrS_2$, i.e. **a**, **b**, **c**, **I** (in the Cr-S plane) and **P** (perpendicular to the Cr-S plane), as shown in the FIG. S5. The easy magnetization axis oscillates between out- (odd number of layers) and in- (even) plane directions. Note that some energy differences are even smaller than 0.01 meV which exceeds the accuracy limit of DFT calculation. For example, the energy difference among directions **b**, **P** and **I** are not distinguished although our convergence criteria are tight enough that the accuracy of total energy was well tested. Nevertheless, all these directions are out-of-plane directions for the $CrS_2$ plane.

| MAE (meV) | axis | 1L | 2L | 3L | 4L |
|---|---|---|---|---|---|
| FM-FM | a | 11.53 | 0.03 | 0.14 | 0 |
| | b | 11.53 | 0 | 0.49 | 0.01 |
| | c | 11.41 | 0.02 | 0 | 0.64 |
| | P | 11.53 | 0.00 | 0.03 | 0.43 |
| | I | 11.52 | 0.00 | 0.03 | 0.01 |
| sAFM-FM | a | 0.12 | 31.06 | 53.92 | |
| | b | 0.11 | 31.00 | 53.97 | |
| | c | 0 | 30.94 | 53.84 | |
| | P | 0.09 | 30.98 | 53.93 | |
| | I | 0.12 | 31.03 | 53.95 | |